\documentclass[journal]{IEEEtran}

\DeclareUnicodeCharacter{03A4}{T}

\ifCLASSINFOpdf
\else
\fi

\usepackage{amsfonts}\usepackage{algorithmic}
\usepackage[ruled,vlined]{algorithm2e}

\usepackage{adjustbox}
\usepackage{graphicx}
\usepackage{caption}
\usepackage[parfill]{parskip}
\usepackage{amsmath}
\usepackage{subcaption} 
\usepackage{placeins} 

\usepackage{booktabs} 
\usepackage{siunitx} 
\usepackage{makecell} 

\usepackage{booktabs}
\usepackage{adjustbox}
\usepackage{multirow}
\usepackage[caption=false,font=normalsize,labelfont=sf,textfont=sf]{subfig}
\usepackage{stfloats}
\usepackage{url}
\usepackage{verbatim}
\usepackage{cite}

\usepackage{booktabs}
\usepackage{multirow}
\usepackage{array}
\usepackage{tabularx}

\usepackage{color,soul}

\usepackage{subcaption}

\begin{document}

\title{SCROOGE: A Physics-Aware Framework for Efficient Orchestration of RIS-Assisted Networks}

\author{\IEEEauthorblockN{
Alexandros I. Papadopoulos,
Sotiris Kopsinos,
Dimitrios Tyrovolas,~\IEEEmembership{Member,~IEEE},
Antonios Lalas, \\
Konstantinos Votis,
George K. Karagiannidis,~\IEEEmembership{Fellow,~IEEE},
and Christos Liaskos 
}
\thanks{A. Papadopoulos and S.Kopsinos are with the Computer Science Engineering Department, University of Ioannina, Ioannina, Greece and with the Information Technologies Institute, CERTH, Greece (e-mails: a.papadopoulos@uoi.gr/alexpap@iti.gr, s.kopsinos@uoi.gr/skopsinos@iti.gr).}
\thanks{A. Lalas, and K. Votis are with the Information Technologies Institute, CERTH, Greece (e-mail: \{lalas, kvotis\}@iti.gr.}
\thanks{D. Tyrovolas and G. K. Karagiannidis are with the Department of Electrical and Computer Engineering, Aristotle University of Thessaloniki, 54124 Thessaloniki, Greece (e-mail: \{tyrovolas, geokarag\}@auth.gr).}
\thanks{C. K. Liaskos is with the Computer Science Engineering Department, University of Ioannina, Ioannina, and with the Foundation for Research and Technology Hellas (FORTH), Greece (e-mail: cliaskos@uoi.gr).}
\thanks{This work was funded by the SNS JU under the EU Horizon Europe research and innovation program through the NATWORK project (Grant No. 101139285).}
}

\maketitle

\begin{abstract}
Reconfigurable Intelligent Surfaces (RISs) are emerging as a key enabler of Programmable Wireless Environments for 6G, but their practical integration into operational networks still lacks orchestration mechanisms that can jointly support resource allocation, energy efficiency, and admission control with low online complexity. This paper presents SCROOGE, a physics-aware orchestration framework for multi-user RIS-assisted networks that operates on information generated offline during RIS codebook compilation, namely optimal codebook entries and per-element influence scores. Rather than relying on online optimization or idealized fading-based abstractions, SCROOGE exploits physics-derived descriptors to support low-latency operating-phase decisions that remain compatible with network-level control requirements. Specifically, SCROOGE introduces: i) an influence-aware, tier-consistent resource-allocation mechanism that combines user priority and element importance in the construction of a common RIS configuration; ii) an energy-efficiency mechanism that deactivates globally low-influence elements; and iii) an admission-control mechanism that accepts or rejects candidate users based on tier-aware compatibility with the currently deployed RIS state, a property that is also relevant when the operator seeks to preserve security-sensitive  conditions. The framework is evaluated through a physics-based electromagnetic setup that captures spatially varying illumination, local mutual coupling, and structured wall-induced multipath. Results show that SCROOGE improves tier-consistency and reduces SNR loss relative to a non-physics-aware baseline, while deactivating a substantial fraction of RIS elements with negligible performance degradation. In addition, the admission-control mechanism sharply separates accepted from rejected users according to tier-specific QoS targets. These findings indicate that physics-aware compilation products can serve as effective enablers of practical RIS orchestration in future 6G networks.

\end{abstract}


%
\IEEEpeerreviewmaketitle

\section{Introduction}

6G networks are expected to support services with stringent latency and reliability requirements, including autonomous driving~\cite{segata2024cooperis}, wireless power transfer~\cite{zhang2018wireless}, and extended reality~\cite{tsimpoukis2024realizing}. Achieving the corresponding Quality of Service (QoS) targets, however, requires mitigating the uncertainty imposed by wireless propagation, which arises from the combined effects of path loss, shadowing, and fading~\cite{liaskos2026tutorialcontrollingmetasurfacesnetwork}. To address this challenge, the concept of Programmable Wireless Environments (PWEs) has emerged, aiming to transform the propagation medium from an uncontrollable factor into a software-defined component of the communication system~\cite{liaskos2026tutorialcontrollingmetasurfacesnetwork}. Under this paradigm, the wireless environment is no longer treated merely as a passive propagation medium, but instead becomes an active participant in the communication process, capable of shaping the channel conditions experienced by wireless links. As a result, PWEs introduce a new degree of freedom in network design, whereby propagation itself can be engineered to improve reliability and stability.

Realizing the PWE paradigm in practice requires physical mechanisms capable of manipulating electromagnetic (EM) wave \cite{lalas2015reconfigurable} propagation in a programmable manner. Among the technologies proposed for this purpose, Reconfigurable Intelligent Surfaces (RISs) have emerged as a promising candidate. A RIS is typically implemented using metamaterial concepts \cite{lalas2014programmable}, whereby a large surface is constructed from repeating unit cells whose EM response can be dynamically tuned~\cite{papadopoulos2022open}. At a macroscopic level, a RIS resembles a thin planar surface composed of many controllable elements whose collective configuration determines the interaction between incident waves and induced surface currents. By programming this configuration, the RIS modifies the interaction between incident waves and the induced surface currents, enabling controlled scattering of EM signals~\cite{ComMag}. Through this mechanism, RISs can implement a variety of macroscopic functionalities, including beam steering and splitting, absorption, and more general wavefront shaping across phase, amplitude, and polarization, as well as sensing-oriented operations~\cite{liaskos2022software}. Thus, when deployed within the wireless environment, RISs can act as controllable propagation elements that complement conventional radio nodes, introducing additional degrees of freedom for network-level control and resource management. This capability supports representative PWE objectives including the restoration of LoS connectivity~\cite{ComMag}, the creation of protected or quiet regions for physical-layer security~\cite{11275068}, environment-assisted localization and sensing~\cite{9293395}, and the mitigation of Doppler effects in V2X systems by manipulating the effective angles of arrival~\cite{basar2021reconfigurable}. As a result, RIS-assisted PWEs gradually shift wireless propagation from a largely uncontrollable entity to an optimizable network resource.

The capabilities offered by RIS-assisted PWEs have also attracted increasing attention from the standardization community, which views programmable propagation as a key enabler of future wireless infrastructures. For instance, the ITU-R IMT-2030 framework identifies advanced propagation control and programmable environments among the technologies expected to support the beyond-5G vision~\cite{ITU_M2160}, while initiatives such as ETSI’s ISG RIS are developing technical reports covering use cases, system architectures, deployment considerations, and channel modeling for RIS-enabled networks~\cite{etsi2023reconfigurable}. These developments indicate that RIS-assisted PWEs are gradually evolving from a purely research concept toward a potential architectural component of next-generation wireless systems. As RIS technologies move closer to practical deployment, their integration into operational wireless networks introduces new requirements for system-level control and coordination. In particular, RIS-assisted environments introduce additional controllable resources at the propagation layer, whose configuration must be coordinated with communication parameters and network-level service requirements. This expanded control space naturally raises questions regarding how such resources should be managed and coordinated within the overall operation of future wireless networks.

\subsection{State-of-the-Art}
Resource management for RIS-assisted networks has been studied from several complementary perspectives, including fairness-oriented allocation, latency-aware control, signaling management, and energy-aware operation. A first line of work formulates joint optimization problems in which the RIS configuration is adapted together with transmit-side variables such as beamforming and power allocation. For example, in~\cite{zhai2024resource}, active RIS-aided multi-cluster SWIPT cooperative NOMA is studied, and beamforming, power allocation, and RIS coefficients are optimized to minimize BS power or maximize the minimum user energy efficiency. Similarly, in~\cite{zhang2022optimal}, a dynamic RIS-assisted wireless network is considered, and online reinforcement learning is used to jointly optimize transmit power and a shared RIS phase-shift matrix with energy efficiency as the objective. A second line of work focuses on delay- and reliability-driven operation. In~\cite{9611292}, RIS-aided eMBB/URLLC traffic multiplexing is studied, and the proactive design of multiple RIS phase-shift matrices at the beginning of each time slot is proposed so that the system can react quickly when URLLC traffic arrives. In~\cite{10636301}, RIS-assisted semi-grant-free NOMA is examined, and the number of admitted low-priority users under delay-outage constraints is maximized through alternating optimization of power and RIS phases. In a related but different direction, Age of Information in RIS-empowered uplink cooperative NOMA is optimized in~\cite{10225434}, again by relying on online optimization of RIS phases and communication variables.


Another relevant line of work examines how RIS resources are shared among users via the partitioning of RIS elements. In~\cite{10978200}, authors propose a low-complexity element allocation strategy for OFDM systems, where RIS elements are assigned one by one to the user that benefits the most in terms of a minimum-rate metric. Moreover, a low-complexity alternative in which RIS elements can be divided among users, alongside a more general framework that jointly optimizes BS beamforming, RIS activation, and phase shifts is investigated in~\cite{he2023reconfigurable}. 


Energy-related aspects have also been studied, but mostly at the level of transmit power, network power minimization, or full-RIS activation decisions. In~\cite{zhai2024resource}, user energy efficiency is defined by accounting for BS power, user hardware power, and the active RIS power budget. In~\cite{zhang2022optimal}, energy efficiency is maximized as the ratio of sum-rate to total power dissipation, including RIS element power consumption. In~\cite{he2023reconfigurable}, green RIS-assisted networks with user admission control are investigated, and the set of active RISs is explicitly optimized to reduce total network power. Admission control has received even less attention in the RIS literature. In~\cite{9611292}, URLLC packet admission is considered jointly with rate-loss control for eMBB users, while in~\cite{10636301} the number of grant-free users that can be supported under delay constraints is determined. In~\cite{he2023reconfigurable}, user admission control is explicitly studied in a multi-RIS downlink, and users are iteratively removed when QoS targets become infeasible. At the network-management level, RIS reconfiguration and handover signaling are analyzed in~\cite{11160619} using stochastic geometry, and admission decisions are discussed through a RIS management entity. 

\subsection{Motivation \& Contributions}

\begin{figure*}[t]
 \centering
  \includegraphics[width=\linewidth]{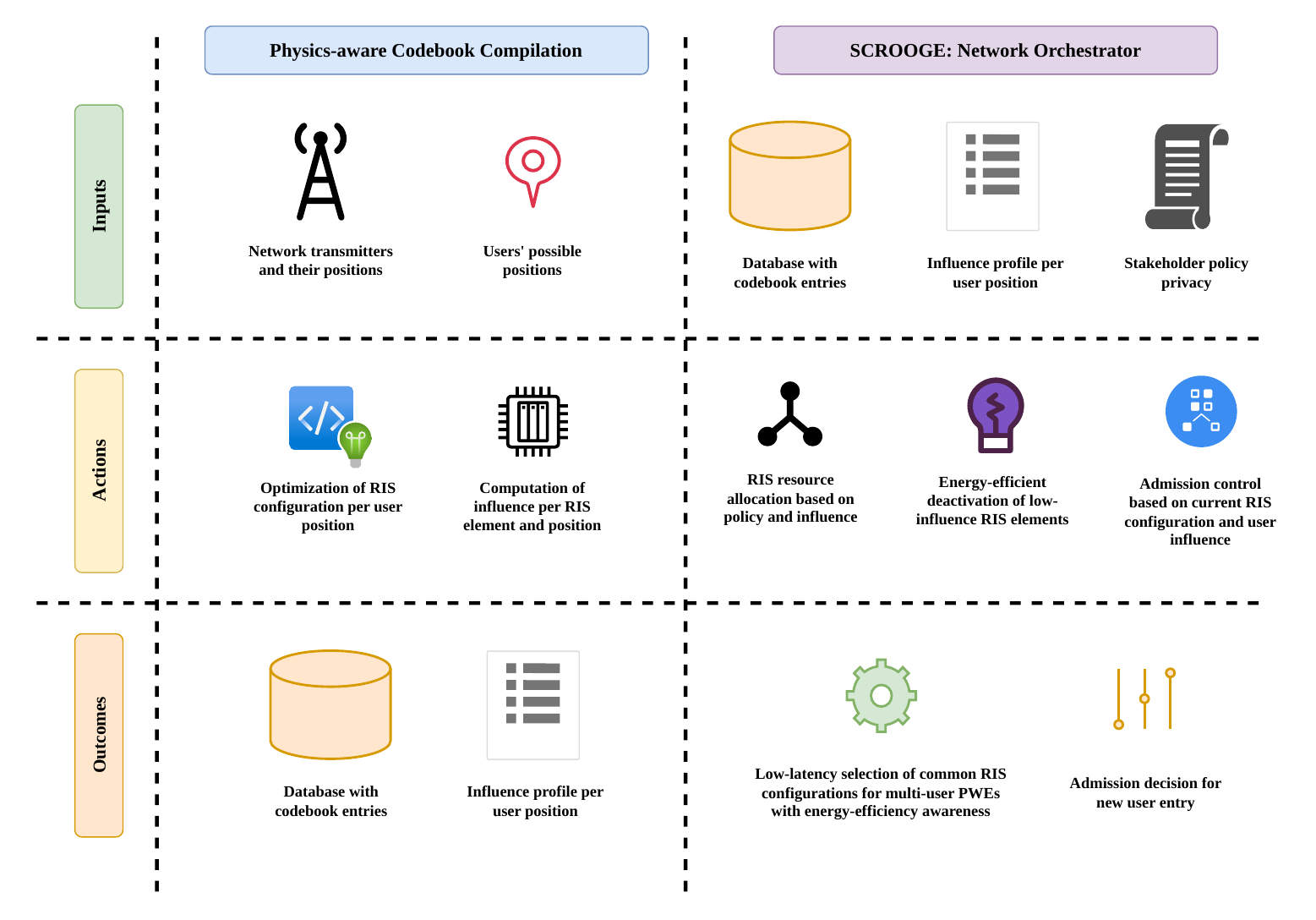}
   \caption{Physics-aware orchestration of RIS-assisted network via SCROOGE framework. Based on users' codebook entries, influence score and operator tier, resource allocation, energy efficiency and admission control are managed.}
\label{fig:system_model}
\end{figure*}

Despite significant progress in RIS-assisted wireless communications, most existing approaches rely on online optimization procedures to determine RIS configurations during network operation. Specifically, while such methods have demonstrated strong performance in controlled scenarios, in practical deployments involving multiple RIS units and a large number of users, repeated optimization introduces additional latency and requires continuous control signaling between RIS units and network entities. This operational model becomes increasingly difficult to sustain under the stringent responsiveness and scalability requirements envisioned for emerging 6G services~\cite{11359620}. One possible way to address these limitations is to separate the computationally intensive RIS synthesis stage from the real-time operation of the network. In such a setting, optimal RIS configurations are determined offline for representative user positions and propagation objectives and are subsequently stored in a codebook that can be accessed during network operation. This two-phase architecture, consisting of an offline synthesis stage and an operating phase where precomputed configurations are selected, is conceptually illustrated in Fig.~\ref{fig:system_model}. In this direction, the work in~\cite{papadopoulos2026physicsawareriscodebookcompilation} demonstrated that incorporating physics-aware descriptors, such as the contribution of individual RIS elements to the resulting macroscopic response, can reduce the effective dimensionality of the synthesis process while improving the quality of the compiled configurations. These descriptors capture how microscopic RIS elements contribute to the resulting propagation functionality and therefore provide additional information beyond the RIS configuration itself. However, while compiled RIS codebooks have primarily been used as lookup tables for selecting configurations, the potential of physics-aware codebook descriptors to support network-level orchestration functions such as resource allocation, energy management, and admission control remains largely unexplored.

In this work, we develop a physics-aware framework named \textit{SCROOGE} that exploits the information embedded in compiled RIS codebooks to support operating-phase orchestration in RIS-assisted networks. In more detail, the proposed approach leverages both the precomputed RIS configurations and the associated per-element influence descriptors obtained during the offline synthesis process, enabling network-level control decisions without requiring repeated online optimization. The main contributions of this paper are summarized as follows:
\begin{itemize}
\item We develop SCROOGE, a physics-aware orchestration framework that exploits compiled RIS codebooks and per-element influence descriptors obtained during the offline synthesis phase in order to support operating-phase control decisions in RIS-assisted networks without requiring repeated online optimization.
\item We design a set of physics-aware orchestration mechanisms that utilize the structural information embedded in compiled codebook entries to support three key network-management functions, namely tier-aware resource allocation, energy-efficient RIS element activation, and admission control under QoS constraints.
\item We evaluate the proposed framework using a physics-based EM simulation environment that explicitly models RIS synthesis and propagation effects, enabling a high-fidelity assessment of the resulting orchestration mechanisms under realistic propagation conditions.
\end{itemize}
\vspace{-3mm}
\subsection{Structure}
The remainder of this paper is organized as follows. Section~\ref{sec:system_model} introduces the system model. Section~\ref{sec:scrooge} presents the proposed SCROOGE orchestration framework and describes its resource allocation, energy-efficiency, and admission-control mechanisms. Section~\ref{sec:performance} evaluates the performance of the proposed framework using a physics-based EM simulation setup. Finally, Section~\ref{sec:conclusion} concludes the paper.

\section{System Model} \label{sec:system_model}

We adopt a physics-based EM setup rather than a purely statistical fading abstraction in order to evaluate orchestration decisions with credible physical grounding. This choice is motivated by the fact that RIS operation is inherently geometry- and configuration-dependent: the link quality at a candidate user location is driven by spatially varying illumination across the aperture, near-field phase curvature, and surface-level interactions that are not faithfully captured by diagonal far-field RIS models under Rayleigh/Rician assumptions. The proposed setup therefore provides a controlled, reproducible testbed where RIS configurations, environment-induced multipath, and resulting Signal-to-Noise Ratio (SNR) maps can be computed directly from first principles and then consumed by the network-layer mechanisms studied in this work.

Specifically, we consider a three-dimensional indoor environment represented by a cubic room of side length $L$. A point transmitter located at $\mathbf{s}\in[0,L]^3$ radiates a continuous-wave signal at carrier frequency $f$, with wavelength $\lambda=c_0/f$ and wavenumber $k=2\pi/\lambda$. RISs are mounted on the vertical walls and manipulate the EM response of the enclosure by programming the phase of the reflected field. Each RIS is modeled as a planar array of $N=N_{\mathrm{d}}\times N_{\mathrm{d}}$ sub-wavelength unit cells placed on a uniform grid with center-to-center spacing $d$, typically chosen such that $d\le \lambda/4$ to enable dense spatial sampling of the impinging field and mitigate spatial aliasing~\cite{9374451}. The $n$-th unit cell is located at geometric center $\mathbf{p}_n=(x_n,y_n,z_n)$ on a wall with outward unit normal $\mathbf{n}_w$, and is characterized by a complex reflection coefficient
\begin{equation}
\Gamma_n=\rho_n e^{j\phi_n},
\end{equation}
where $\phi_n$ is electronically controlled and $\rho_n$ denotes the reflection amplitude (with $\rho_n=1$ for purely passive operation).

To capture spatial non-uniformity over the surface and environmental interactions, we explicitly evaluate the incident field at each unit cell~\cite{11275068}. In particular, the field impinging on cell $n$ is decomposed into (i) direct illumination from the transmitter, (ii) re-illumination due to EM interactions with nearby unit cells (mutual coupling), and (iii) secondary contributions induced by specular reflections from the uncovered portions of the surrounding walls. Accordingly, we write
\begin{equation}
E_{\mathrm{inc},n}=E_{\mathrm{dir},n}+E_{\mathrm{cpl},n}+\sum_{w\in\mathcal{W}}E_{\mathrm{sec},n}^{(w)},
\label{eq:einc}
\end{equation}
where $\mathcal{W}$ denotes the set of reflective walls not covered by RIS.

Let $\mathbf{d}_n=\mathbf{p}_n-\mathbf{s}$, $r_n=\|\mathbf{d}_n\|$, and $\hat{\mathbf{d}}_n=\mathbf{d}_n/r_n$. The direct contribution at cell $n$ is modeled as a spherical wave with optional angular weighting that captures a directive transmitter pattern and an element-dependent cosine response~\cite{pitilakis2023mobility}: 
\begin{equation}
E_{\mathrm{dir},n}
=\frac{e^{jkr_n}}{r_n}\,
\bigl|\hat{\mathbf{d}}_n\cdot\mathbf{u}_{\mathrm{TX}}\bigr|^{m}\,
\Bigl[\max\!\bigl\{0,\hat{\mathbf{d}}_n\cdot\mathbf{n}_w\bigr\}\Bigr]^{p},
\label{eq:edir}
\end{equation}
where $\mathbf{u}_{\mathrm{TX}}$ denotes the transmitter main-lobe direction, $m\ge 0$ controls the beamwidth of the radiation pattern, and $p\ge 0$ tunes the angular sensitivity of each unit cell (larger $p$ yields a more specular response).

The mutual coupling effect is approximated by dominant interactions within a local neighborhood $\mathcal{N}_n$ (e.g., the four or eight nearest neighbors on the grid). The coupling-induced field at cell $n$ is modeled as
\begin{equation}
E_{\mathrm{cpl},n}
=\alpha\sum_{m\in\mathcal{N}_n}\Gamma_m\,E_{\mathrm{inc},m}\,
\frac{e^{jk\|\mathbf{p}_n-\mathbf{p}_m\|}}{\|\mathbf{p}_n-\mathbf{p}_m\|},
\label{eq:ecpl}
\end{equation}
where $\alpha\in[0,1]$ is a dimensionless coefficient controlling the coupling strength. Note that Eq. \eqref{eq:ecpl} depends on $\{E_{\mathrm{inc},m}\}$, which implies that Eq. \eqref{eq:einc} constitutes a coupled system over the RIS grid.

To incorporate additional illumination paths produced by specular reflections from uncovered walls, we adopt an image-source model. For each wall $w\in\mathcal{W}$, we place an image of the transmitter at $\mathbf{s}_w$ and weight its contribution by the wall reflectivity $\beta_w$. The resulting secondary term at cell $n$ is expressed as
\begin{equation}
E_{\mathrm{sec},n}^{(w)}
=\beta_w\frac{e^{jk\|\mathbf{p}_n-\mathbf{s}_w\|}}{\|\mathbf{p}_n-\mathbf{s}_w\|}\,
\Bigl[\max\!\bigl\{0,\hat{\mathbf{d}}_{n,w}\cdot\mathbf{n}_w\bigr\}\Bigr]^{p},
\label{eq:esec}
\end{equation}
where $\hat{\mathbf{d}}_{n,w}=(\mathbf{p}_n-\mathbf{s}_w)/\|\mathbf{p}_n-\mathbf{s}_w\|$. When an angular dependence is not required, the cosine-law factor in Eq. \eqref{eq:esec} can be omitted, recovering a purely geometric image-source contribution. Given the incident field samples $\{E_{\mathrm{inc},n}\}$ and the programmable coefficients $\{\Gamma_n\}$, the total electric field at an observation point $\mathbf{r}\in[0,L]^3$ is computed via coherent superposition of the reradiated contributions from all unit cells:
\begin{equation}
E(\mathbf{r})
=\sum_{n=1}^{N}\Gamma_n\,E_{\mathrm{inc},n}\,
\frac{e^{jk\|\mathbf{r}-\mathbf{p}_n\|}}{\|\mathbf{r}-\mathbf{p}_n\|}.
\label{eq:efield}
\end{equation}
The received signal quality at $\mathbf{r}$ is quantified through the SNR under normalized noise power, defined as
\begin{equation}
\mathrm{SNR}(\mathbf{r})
=10\log_{10}\!\Bigl(\bigl|E(\mathbf{r})\bigr|^2\Bigr).
\label{eq:snr}
\end{equation}
This formulation retains the core physics relevant to near-field focusing in enclosed spaces: spatially varying illumination across the RIS aperture, local EM interactions among adjacent unit cells, and structured multipath induced by specular wall reflections, while enabling direct evaluation of SNR at arbitrary locations within the room. Importantly, the role of this EM setup is not to re-open physical-layer design questions, but to provide high-fidelity inputs and evaluation for the orchestration layer. The EM model is used only to generate realistic, position-dependent performance surfaces and to stress-test the proposed framework under structured propagation effects typical of indoor deployments. As such, it serves as an enabling instrument that avoids overly idealized channel assumptions and mitigates the risk of conclusions that are artifacts of simplified fading models.

\section{SCROOGE: A Physics-aware Framework for Resource Allocation, Energy Efficiency \& Admission Control}\label{sec:scrooge}

In this section we present SCROOGE, a framework designed to orchestrate RIS-assisted networks during system operation by jointly supporting resource allocation, energy efficiency, and admission control decisions. As illustrated in Fig.~\ref{fig:system_model}, SCROOGE operates on information generated during a physics based codebook compilation process performed offline. This process associates each user location with an optimal codebook entry that describes the RIS configuration required to achieve the desired propagation response. In addition, it provides an influence score for every RIS element that quantifies the contribution of that element to the resulting macroscopic RIS behavior. During network operation multiple users may request service simultaneously and each of them is, therefore, associated with a different RIS configuration stored in the compiled codebook. At the same time SCROOGE receives the tier classification assigned to each user according to the pricing policy defined by the telecommunication stakeholder. The orchestration task of the framework is therefore to determine a common RIS configuration denoted by $CC$ that can simultaneously serve the active users while preserving both the physical consistency encoded in the compiled codebook entries $CE$ and the priority structure implied by the user tiers.

\begin{algorithm}[t]
\caption{SCROOGE: Resource Allocation}
\label{alg:scrooge-alloc}
\KwIn{Codebook entries $\{CE_1,\dots,CE_K\}$, influence profiles $\{v_1,\dots,v_K\}$, payment factors $\{\mathrm{PF}_1,\dots,\mathrm{PF}_K\}$, thresholds $\tau_{\text{low}},\tau_{\text{high}}$, exponents $\alpha,\beta$, small constant $\varepsilon_{\mathrm{inf}}$, quantization level $N_d$}
\KwOut{Common RIS configuration $CC$}

\ForEach{element $n$}{
  $\textit{maxV}_n \leftarrow \max_k\, v_{k,n}$\;
  \uIf{$\textit{maxV}_n \le \tau_{\text{low}}$}{
    $\eta_n \leftarrow 0$\;
  }
  \uElseIf{$\textit{maxV}_n \ge \tau_{\text{high}}$}{
    $\eta_n \leftarrow 1$\;
  }
  \Else{
    $\eta_n \leftarrow \dfrac{\textit{maxV}_n - \tau_{\text{low}}}{\tau_{\text{high}} - \tau_{\text{low}}}$\;
  }
}

\ForEach{phase state $s \in \{1,\dots,N_d\}$ and element $n$}{
  $\textit{scores}(s,n) \leftarrow 0$\;
}

\For{$k \leftarrow 1$ \KwTo $K$}{
  $DE_k \leftarrow \mathrm{Quantize}(CE_k, N_d)$\;
  \ForEach{element $n$}{
    Let $s_{k,n}$ be the phase state of $DE_k$ at element $n$\;
    $W_{k,n} \leftarrow (1 - \eta_n)\,\mathrm{PF}_k^{\alpha} + \eta_n\,\mathrm{PF}_k^{\alpha}(\varepsilon_{\mathrm{inf}} + v_{k,n})^{\beta}$\;
    $\textit{scores}(s_{k,n}, n) \leftarrow \textit{scores}(s_{k,n}, n) + W_{k,n}$\;
  }
}

\ForEach{element $n$}{
  $CC(n) \leftarrow \arg\max_{s \in \{1,\dots,N_d\}} \textit{scores}(s,n)$\;
}

\Return $CC$
\end{algorithm}

Since a single RIS surface can realize only one configuration at a given time the configurations associated with multiple users cannot be applied independently. Practical RIS hardware further restricts the available configurations because each element can realize only a finite set of reflection phase states. Consequently every candidate configuration must be expressed in terms of the discrete phase levels supported by the surface according to the available quantization level $N_d$. Under these constraints the orchestration problem reduces to constructing a configuration that reconciles the individual user configurations while respecting the service priorities defined by the network operator. For this purpose SCROOGE retrieves from the codebook database the entries corresponding to the users currently participating in the network and maps each entry to the discrete phase levels supported by the RIS~\cite{TNSM_RIS_NETWORK}. The tier assigned to each user is then translated into an integer voting weight through a payment factor denoted by $\mathrm{PF}_k$. The common RIS configuration is obtained by applying a weighted majority decision independently at every RIS element across the discretized user entries. Through this procedure the resulting configuration multiplexes the single user optimal solutions contained in the codebook while preserving the hierarchy imposed by the user tiers.

\begin{algorithm}[t]
\caption{SCROOGE: Energy‑Efficiency}
\label{alg:scrooge-off}
\KwIn{Common configuration $CC$ (phase values), influence profiles $\{v_1,\dots,v_K\}$, OFF threshold $\tau_{\text{off}}$}
\KwOut{Energy‑efficient configuration $\widetilde{CC}$}
$\widetilde{CC} \leftarrow CC$\;
\ForEach{element $n$}{
  $\textit{maxV}_n \leftarrow \max\, v_{k,n}$\;
  \If{$\textit{maxV}_n < \tau_{\text{off}}$}{
    $\widetilde{CC}(n) \leftarrow 0$ \tcp{Switch element off}
  }
}
\Return $\widetilde{CC}$
\end{algorithm}

The aggregation mechanism described above relies exclusively on the tier based voting weights assigned to the users and therefore treats all RIS elements in the same manner. Simultaneously, the physics based codebook compilation process provides additional information about the relative importance of individual RIS elements through the influence scores associated with each user configuration. These scores reveal that the contribution of RIS elements to the resulting macroscopic response is not uniform and that some elements may have a stronger impact on the achievable user performance than others. SCROOGE therefore incorporates this information into the aggregation process so that the configuration decision can account not only for the user tiers but also for the physical relevance of each RIS element. To achieve this objective the influence vector associated with each user is first normalized and smoothed in order to obtain a stable measure of the relative importance of every RIS element. For each element $n$ a blending factor denoted by $\eta_n$ and taking values in the interval $[0,1]$ is then computed according to the maximum influence observed across the active users together with two predefined thresholds $\tau_{\text{low}}$ and $\tau_{\text{high}}$. When the influence of an element remains below the lower threshold the configuration decision relies exclusively on the tier based voting weights. When the influence exceeds the upper threshold the voting weights incorporate both the payment factor $\mathrm{PF}$ and the influence scores. For intermediate influence levels the two regimes are combined through linear blending so that the decision progressively shifts from purely tier driven aggregation to influence aware aggregation. 

The resulting aggregation rule can be expressed through the element wise weight assigned to user $k$ at RIS element $n$, which is given by
\[
W_{k,n} = (1-\eta_n)\,\mathrm{PF}_k^{\alpha} + \eta_n\,\mathrm{PF}_k^{\alpha}\,(\varepsilon_{\mathrm{inf}} + v_{k,n})^{\beta},
\]
where $v_{k,n}$ denotes the smoothed influence value associated with the $n$-th element and user $k$, $\mathrm{PF}_k$ is the voting parameter determined by the user tier, $\alpha$ controls how strongly the tier hierarchy affects the aggregation, $\beta$ determines the sensitivity of the decision to the influence scores, and $\varepsilon_{\mathrm{inf}}$ ensures numerical stability when the influence values are very small. The detailed workflow of the physics aware resource allocation mechanism is summarized in Alg.~\ref{alg:scrooge-alloc}. For every RIS element the algorithm evaluates the cumulative score associated with each candidate phase state and selects the state that maximizes the aggregated weight across the participating users. By adjusting the parameters $\alpha$, $\beta$, and the thresholds $(\tau_{\text{low}},\tau_{\text{high}})$ the network operator can regulate the balance between strict tier prioritization and the exploitation of RIS elements that are most beneficial for the users currently present in the network.

Once the common RIS configuration $CC$ has been determined through the resource allocation mechanism the influence information available from the codebook also allows the framework to regulate the activation of RIS elements. The influence scores reveal that certain elements may have negligible impact on the performance of all participating users. In such cases maintaining these elements in an active state does not contribute meaningfully to the resulting RIS response while still consuming hardware resources. SCROOGE therefore employs a physics aware energy efficiency mechanism that selectively deactivates elements whose contribution remains insignificant. The workflow of this mechanism is summarized in Alg.~\ref{alg:scrooge-off}. For each RIS element $n$ the framework evaluates $\textit{maxV}_n$, which corresponds to the maximum smoothed influence of that element across the participating users. This value is then compared to a predefined threshold $\tau_{\text{off}}$. When $\textit{maxV}_n$ falls below the threshold the element is switched off by assigning it a zero state. Otherwise the phase state determined by the resource allocation mechanism is retained. Through this procedure the RIS configuration remains consistent with the physical requirements of the active users while enabling selective element deactivation and reducing the overall energy consumption of the surface. In addition, by suppressing globally non-critical elements, the mechanism avoids maintaining unnecessary scattering degrees of freedom, which may be useful in more security-aware operation where the operator prefers tighter control over how the surface perturbs the surrounding field.

\begin{algorithm}[t]
\caption{SCROOGE: Admission Control}
\label{alg:scrooge_admission}
\KwIn{Current common configuration $CC$, candidate optimal codeword $CE$, candidate influence profile $v_{\text{cand}}$, thresholds $y$ and $x$ for the tier of candidate user.}
\KwOut{Admission decision $\mathsf{admit}\in\{0,1\}$}
\BlankLine

\ForEach{$n \in \mathcal{T}$}{
  $\phi_{\text{common}} \leftarrow CC(n)$\;
  $\phi_{\text{opt}} \leftarrow CE(n)$\;
  $\Delta\phi_n \leftarrow \left|\mathrm{wrap}_{[-\pi,\pi)}\!\left(\phi_{\text{common}}-\phi_{\text{opt}}\right)\right|$\;
  \If{$\Delta\phi_n \le x\,2\pi$}{
    $c \leftarrow c + 1$\;
  }
}

\If{$c \ge \left\lceil y\,|\mathcal{T}| \right\rceil$}{
  $\mathsf{admit} \leftarrow 1$\;
}
\Else{
  $\mathsf{admit} \leftarrow 0$\;
}
\Return $\mathsf{admit}$
\end{algorithm}

 \begin{figure}[t]
 \centering
  \includegraphics[width=\linewidth]{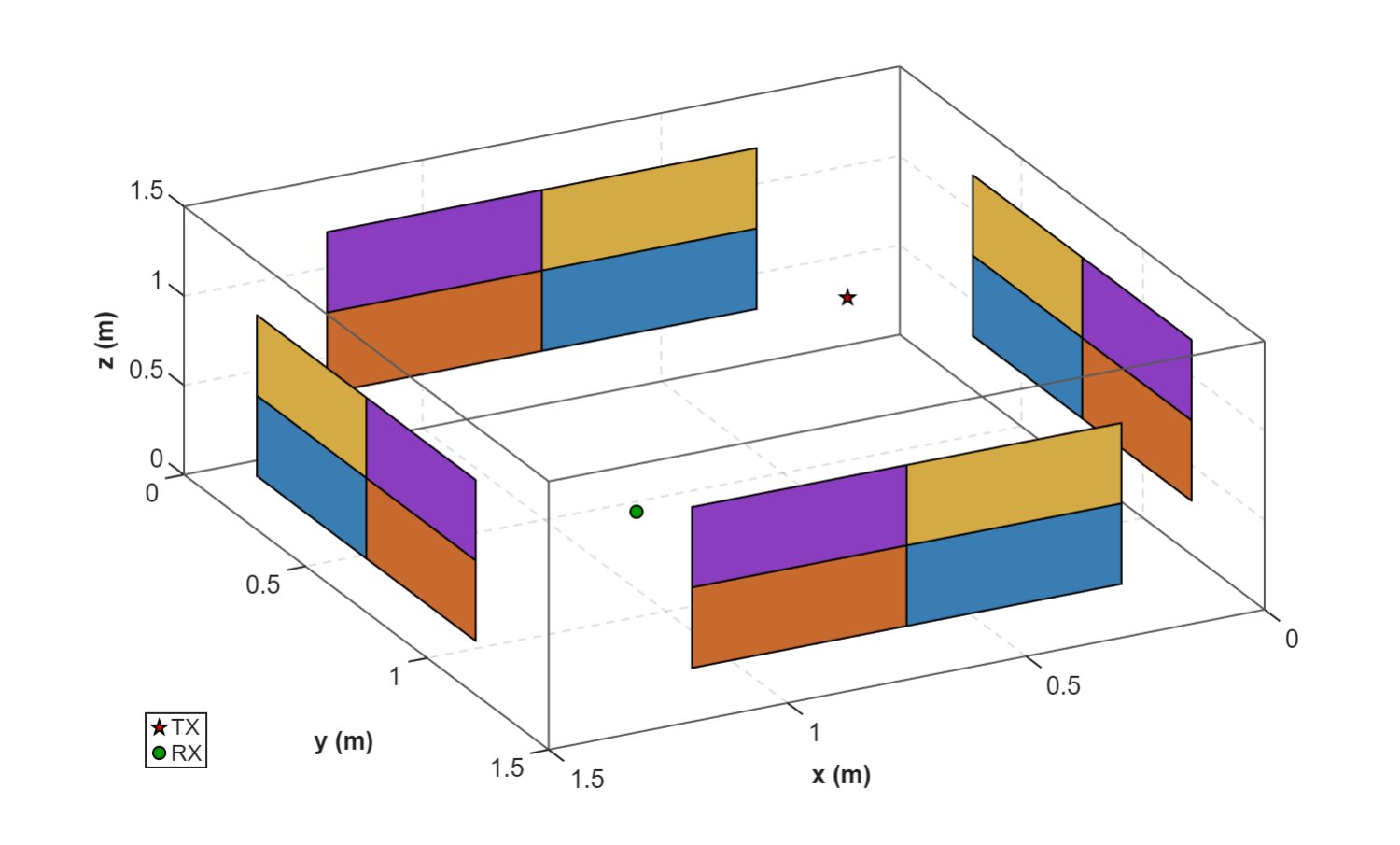}
    \caption{Illustration of the simulated, physics-based EM setup.}
 \label{fig:setup}
\end{figure}

The same physics aware information also supports admission control when additional users request service. At the moment a candidate user arrives the network already serves $K_{\text{current}}$ users and the corresponding common configuration $CC$ has been established through the resource allocation and energy efficiency mechanisms. To evaluate the compatibility of the candidate user with the current configuration SCROOGE retrieves the candidate codebook entry $CE$, the associated influence scores, and the tier information represented by the payment factor $\mathrm{PF}$. The algorithm then identifies the subset of RIS elements whose influence for the candidate user is among the largest values. This subset, denoted by $\mathcal{T}$, corresponds to the top $y\%$ of elements ranked according to their influence. For these elements the phase difference between the currently deployed configuration $CC$ and the candidate configuration $CE$ is evaluated. Admission is granted when at least $y\%$ of the elements in $\mathcal{T}$ exhibit a phase mismatch that does not exceed a tolerance corresponding to $x$ of $2\pi$. The tolerance parameter is defined in a tier dependent manner so that higher tier users require stricter phase compatibility while lower tier users may be admitted under more relaxed conditions. The operator may also assign different values of $y$ across tiers in order to implement more flexible admission policies. More broadly, this mechanism can also be viewed as preserving the currently deployed propagation state, since a new user is admitted only when its most critical RIS elements can be accommodated without causing excessive disruption, a property that is also relevant under security-aware or protected-region constraints. The workflow of this mechanism is summarized in Alg.~\ref{alg:scrooge_admission}.

Taken together these mechanisms establish SCROOGE as an operating phase orchestration framework that combines economic prioritization with the physics derived information extracted during codebook compilation. The framework operates entirely at the orchestration level by consuming the compiled codebook descriptors and directly determining the deployed RIS configuration without requiring iterative control loops between the RIS and the users or any online optimization procedures. As a result the signaling overhead remains low and the additional decision latency is minimal since all decisions rely on precomputed information combined with lightweight rule based aggregation. This design enables SCROOGE to support multi user RIS networks in which fairness, performance, and energy consumption must be managed jointly under practical hardware constraints and emerging 6G system requirements.

\begin{figure*}[t]
 \centering
  \includegraphics[width=\linewidth]{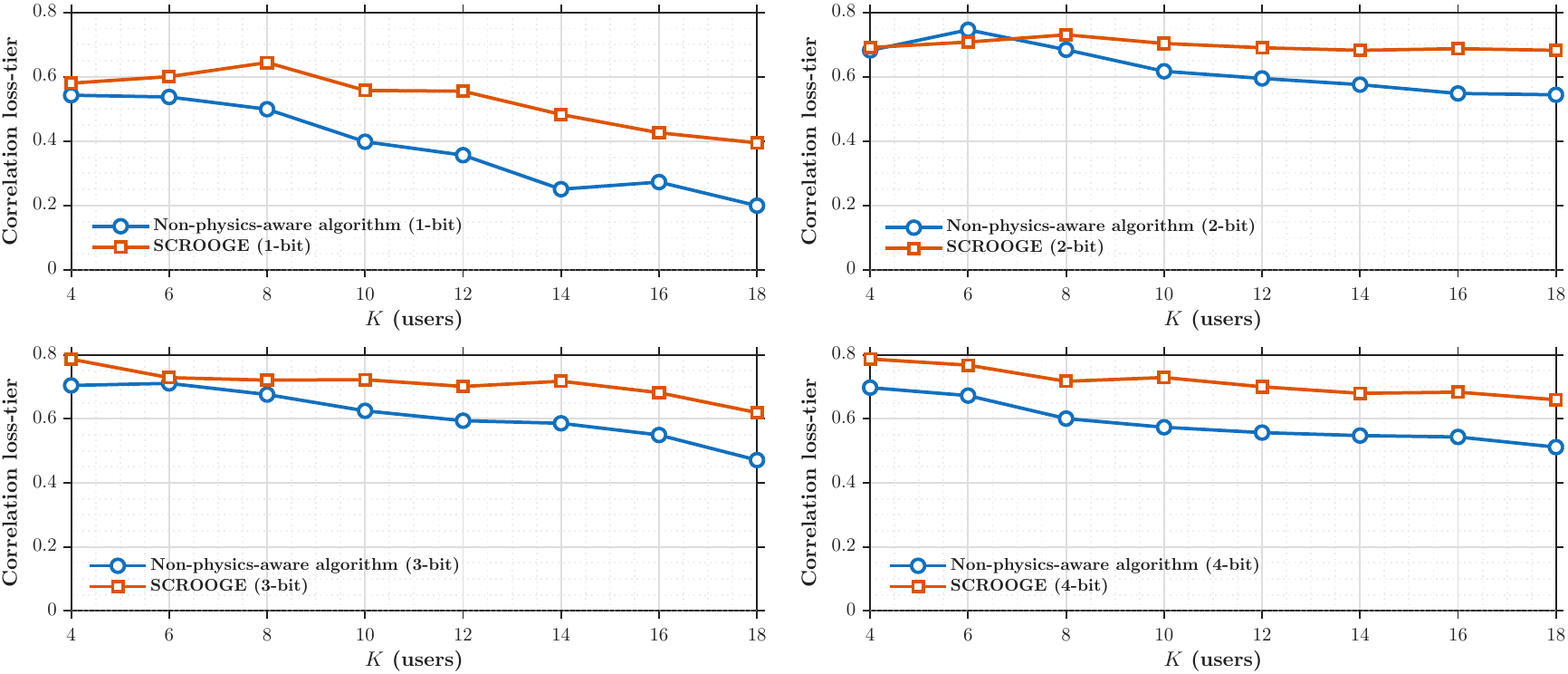}
    \caption{Correlation between users' tiers and SNR loss of SCROOGE compared with the non-physics-aware baseline.}
 \label{fig:SCROOGE_fairness}
\end{figure*}

\begin{figure*}[t]
 \centering
  \includegraphics[width=\linewidth]{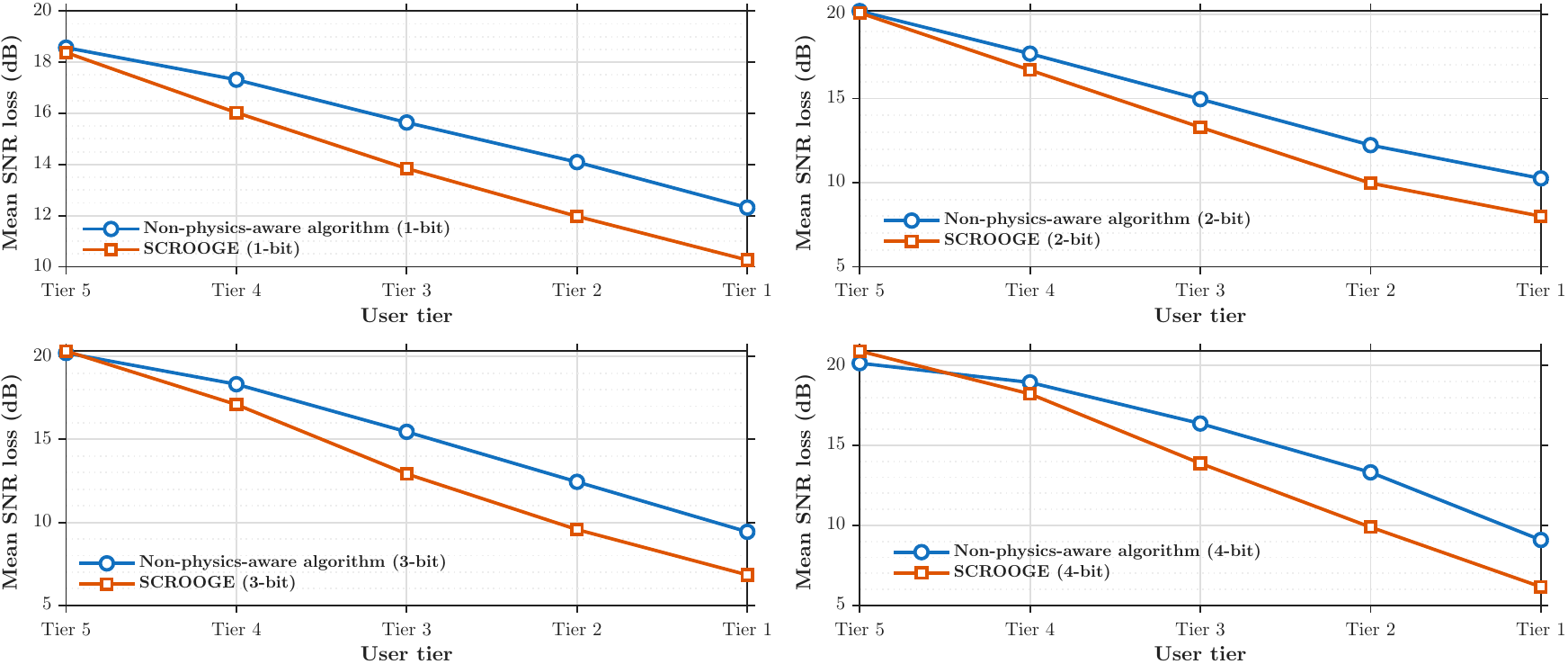}
    \caption{SNR loss across all the tiers of SCROOGE compared with the non-physics-aware baseline.}
 \label{fig:SCROOGE_SNRLoss}
\end{figure*}

\section{Performance Evaluation}\label{sec:performance}

In this section, we present the evaluation results of each SCROOGE mechanism. All experiments are conducted using the physics-based EM setup described in Section~\ref{sec:system_model}, with the corresponding parameter values summarized in Table~\ref{tab:emulation_setup_values}. The setup is illustrated in Fig.~\ref{fig:setup}. In each Monte Carlo realization, user locations are randomly selected from the considered deployment area, so that the reported results reflect performance under diverse spatial user configurations. Throughout the evaluation, Tier 1 denotes the highest-priority service class, while Tier 5 denotes the lowest-priority class.

\begin{table}[!t]
\caption{Simulation Parameters}
\label{tab:emulation_setup_values}
\centering
\renewcommand{\arraystretch}{1.05}
\begin{tabularx}{\columnwidth}{|l|X|}
\hline
\textbf{Parameter} & \textbf{Value} \\
\hline
Environment size & $L=1.5\,\text{m}$ \\
\hline
Frequency / wavelength & $f=6\,\text{GHz}\ \Rightarrow\ \lambda = c_0/f \approx 0.05\,\text{m}$ \\
\hline
RIS layout & $N_r=N_c=60,\quad d=\lambda/4\ \ (\text{with } d\le \lambda/4)$ \\
\hline
Walls covered \& RIS elements & $N_{\text{walls}}=4,\quad N_{total} = 4\times N_r \times N_c \times N_{\text{walls}}=57{,}600$ \\
\hline
Transmitter position & $\mathbf{s} = [0.2L,\,0.25L,\,0.5L]$ \\
\hline
Mutual coupling strength & $\alpha=0.15$ \\
\hline
\end{tabularx}
\end{table}

Specifically, we construct a codebook database that stores, for each candidate user location, (i) the corresponding single-user optimal RIS configuration, (ii) the associated per-element influence score, and (iii) the achieved single-user optimal SNR, which is retained as a benchmarking reference. We assume that the telecommunication operator classifies users into five tiers and we evaluate SCROOGE via $2{,}000$ Monte Carlo realizations by randomly selecting the active users from the database and randomly assigning a tier to each selected user in every realization. For the resource-allocation mechanism, we map the voting parameters to $\mathrm{PF}_1–\mathrm{PF}_5\in\{5,4,3,2,1\}$, set $\tau_{\text{low}}=0.3$ and $\tau_{\text{high}}=0.8$, choose $\alpha=1$ and $\beta=1.5$, and use $\varepsilon_{\mathrm{inf}}=10^{-3}$, while testing quantization resolutions of 1, 2, 3, and 4 bits. For the energy-efficiency mechanism, we set $\tau_{\text{off}}=0.25$. For admission control, we use tier-dependent phase-mismatch tolerances $x=\{60,45,30,25,15\}\%$ of $2\pi$ from the lowest to the highest tier, and we fix the top-influence fraction to $y=10\%$ for all tiers. 

\subsection{Physics-aware Resource Allocation}

First, we assess the fairness gains enabled by SCROOGE's physics-driven resource allocation compared with the non-physics baseline. The results are shown in Fig.~\ref{fig:SCROOGE_fairness}, where tier-consistency is quantified via the correlation between each user's tier and the corresponding SNR loss for different numbers of active users $K$. Across all quantization levels, SCROOGE consistently achieves higher correlation than the non-physics baseline, indicating that influence-aware voting better preserves the tier hierarchy under multiplexing. The improvement is most pronounced under coarse control: in the $1$-bit case, the non-physics baseline degrades from $0.54$ at $K=4$ to $0.20$ at $K=18$, whereas SCROOGE decreases more gracefully from $0.58$ to $0.39$. At intermediate loads, the gap remains substantial. For example, at $K=10$ and $K=12$, the non-physics baseline attains $0.39$ and $0.38$, while SCROOGE achieves $0.56$ and $0.55$, respectively. The same behavior appears at moderate resolutions: in the $2$-bit case, the non-physics baseline attains a mean correlation of $0.62$ with standard deviation $0.07$, whereas SCROOGE increases the mean to $0.69$ with a much smaller standard deviation of $0.016$. Similarly, in the $3$-bit case, the non-physics baseline yields $0.61 \pm 0.08$ while SCROOGE reaches $0.71 \pm 0.04$. Even with finer quantization, the trend persists: for $4$-bit, the non-physics baseline spans approximately $0.70 \rightarrow 0.51$, while SCROOGE remains higher at $0.79 \rightarrow 0.66$. Overall, these results match the SCROOGE design principle: when element influence is high, the voting weights shift from purely tier-based to tier-plus-physics weighting, which mitigates tier inversions that arise when the non-physics baseline treats all RIS elements as equally important.

\begin{figure}[t]
 \centering
  \includegraphics[width=\linewidth]{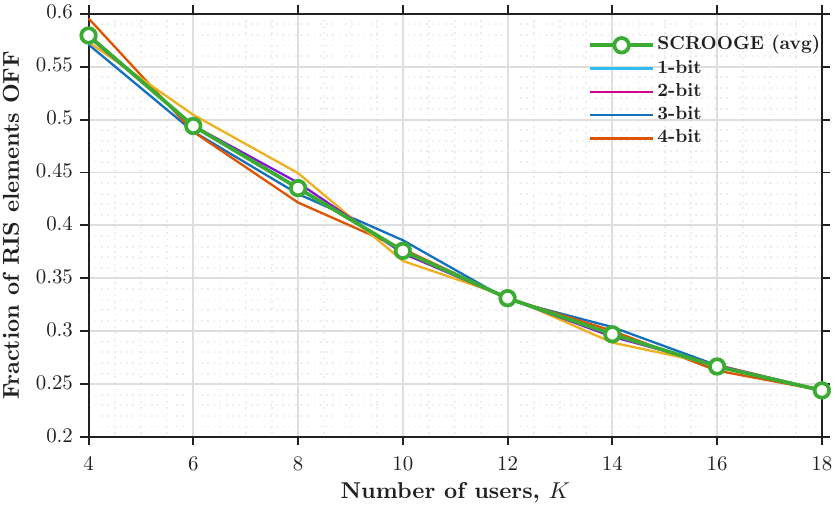}
    \caption{Fraction of the RIS elements that are deactivated using SCROOGE in different network loads.}
 \label{fig:SCROOGE_EE}
\end{figure}

\begin{figure}[t]
 \centering
  \includegraphics[width=\linewidth]{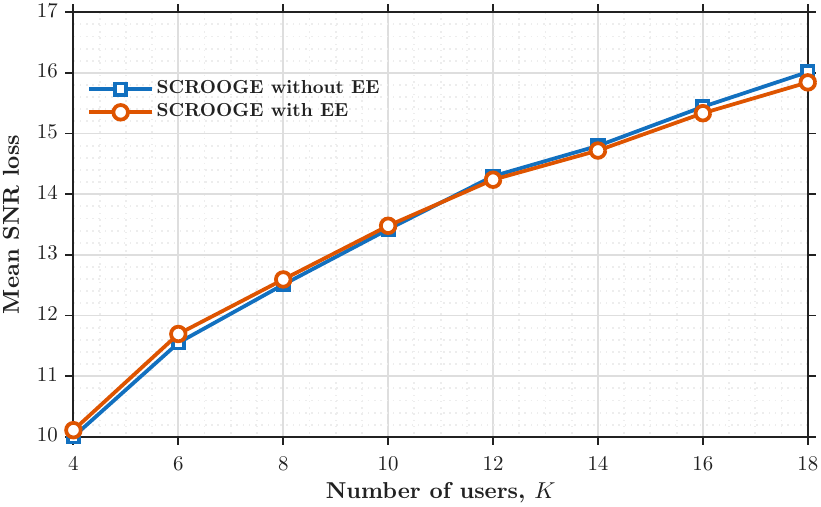}
    \caption{Comparison of SNR with and without SCROOGE energy efficiency usage.}
 \label{fig:SCROOGE_with_withoutEE}
\end{figure}

Beyond fairness, we also examine whether users benefit in terms of absolute performance, quantified by the reduction of the SNR loss relative to each user's single-user optimum stored during codebook compilation. Fig.~\ref{fig:SCROOGE_SNRLoss} reports the mean SNR loss per tier, aggregated over all user-count cases and Monte Carlo trials (the same $2{,}000$ realizations as before), and therefore reveals how the multiplexing cost is distributed across tiers. The non-physics baseline follows a predominantly price-driven dominance, which tends to penalize lower-tier users on RIS elements that are in fact critical for their channels. Addressing this, SCROOGE alleviates this effect by prioritizing influential elements while still respecting the tier hierarchy.
This improvement is already visible under coarse quantization. In the $1$-bit case, the non-physics baseline yields tier-wise losses of $18.56$, $17.31$, $15.64$, $14.09$, and $12.31$~dB, whereas SCROOGE reduces them to $18.38$, $16.03$, $13.84$, $11.97$, and $10.26$~dB, demonstrating gains across all tiers. The same trend holds for $2$-bit, where the non-physics baseline reports $20.20$, $17.66$, $14.96$, $12.23$, and $10.25$~dB and SCROOGE improves these to $20.09$, $16.69$, $13.29$, $9.97$, and $8.00$~dB. As the hardware constraints become less restrictive, the benefits become clearer, especially for the higher tiers: in the $4$-bit case the higher-tier endpoints improve from $13.31 \rightarrow 9.89$~dB and $9.09 \rightarrow 6.16$~dB, corresponding to enhancements of approximately $3.4$~dB and $3$~dB, respectively. Similarly, in the $3$-bit case the highest-tier endpoint decreases from $9.43 \rightarrow 6.84$~dB. Overall, these results support the conclusion that influence-aware aggregation reduces the systematic disadvantage of all the tiers under a single common configuration, improving service balance without violating the intended tier ordering. A further qualitative observation is that, under SCROOGE, the tier-loss curve becomes closer to linear, which is consistent with the stronger tier-consistency correlations reported in Fig.~\ref{fig:SCROOGE_fairness}.

\begin{figure}[t]
 \centering
  \includegraphics[width=\linewidth]{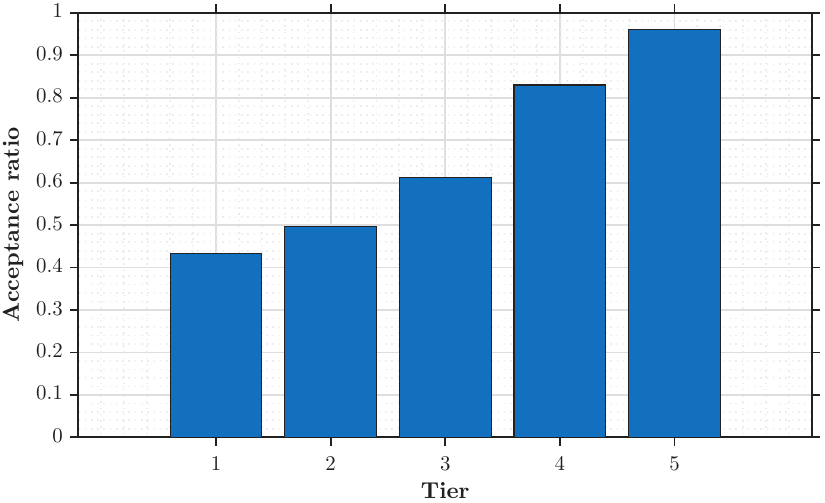}
    \caption{Acceptance ratio per users' tier using SCROOGE admission control.}
 \label{fig:AcceptanceRatio_AC}
\end{figure}

\begin{figure*}[t]
 \centering
  \includegraphics[width=\linewidth]{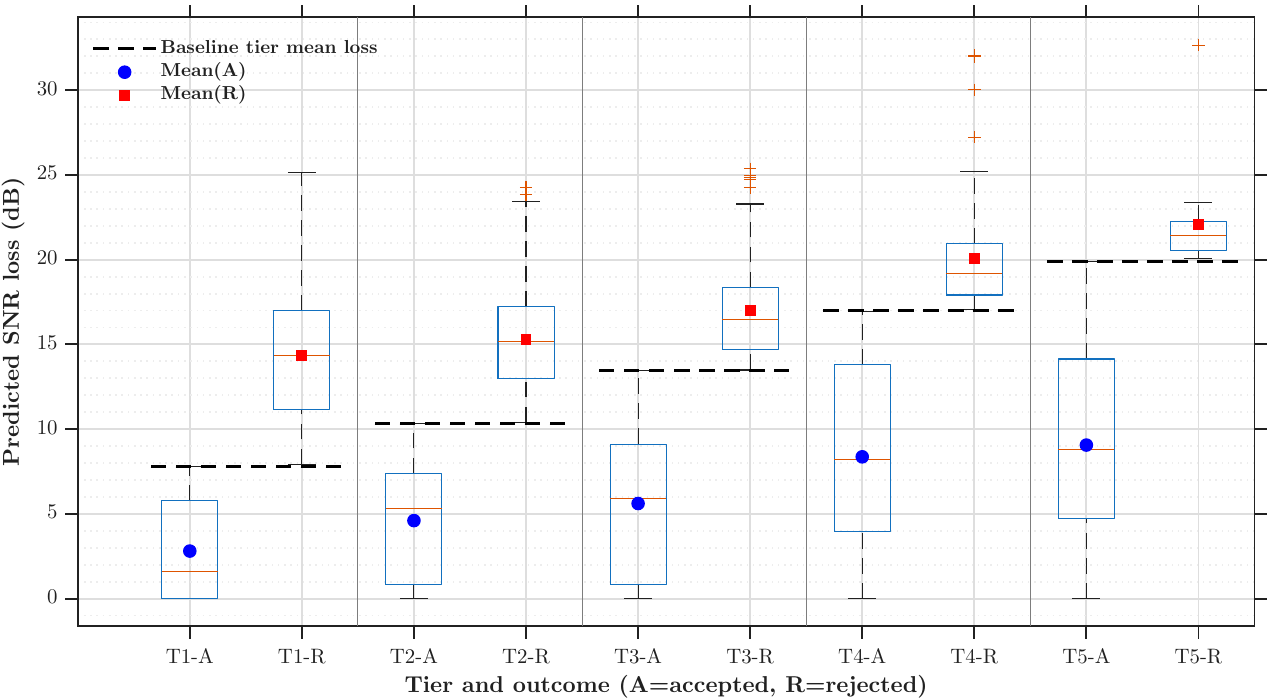}
    \caption{SNR loss statistics per users' tier for accepted and rejected user using SCROOGE admission control mechanism.}
 \label{fig:SNRLoss_AC}
\end{figure*}

\subsection{Physics-aware Energy Efficiency}

Fig.~\ref{fig:SCROOGE_EE} evaluates SCROOGE's energy-efficiency mechanism through the mean OFF fraction, i.e., the proportion of RIS elements switched off because they are weakly influential for all users. The dominant trend is load dependence: as the number of active users $K$ increases, more RIS elements become influential for at least one user, and therefore fewer elements can be safely deactivated. When averaged over all quantization levels, the OFF fraction decreases from approximately $0.58$ at $K=4$ to approximately $0.24$ at $K=18$, indicating that even under heavy load about $24\%$ of the RIS can still be turned off. The corresponding intermediate averaged values for $K=6$ to $K=16$ are $0.49$, $0.43$, $0.37$, $0.33$, $0.30$, and $0.27$, confirming a smooth reduction as the network becomes more populated. Importantly, the OFF curves remain very similar across $1$--$4$ bits (e.g., at $K=4$ they lie around $0.57$--$0.60$, and at $K=16$ around $0.26$--$0.27$), which indicates that the OFF decision is driven primarily by the physics-derived influence structure and the network load, rather than by phase resolution. This behavior matches the intent of SCROOGE: energy control is a lightweight operating-phase action that exploits compilation-time influence information to deactivate elements that contribute negligibly to all users.

Subsequently, we investigate how the deactivation mechanism affects the performance of the users. 
Fig.~\ref{fig:SCROOGE_with_withoutEE} validates that the OFF rule achieves energy savings with negligible performance degradation by comparing SCROOGE with OFF enabled versus disabled.
The curves remain extremely close across all $K$, showing that switching off globally low-influence elements removes mostly non-contributing degrees of freedom.
Quantitatively, at $K=4$ the mean loss is $10.11\,\text{dB}$ (OFF enabled) versus $10.00\,\text{dB}$ (OFF disabled) ($\approx 0.11\,\text{dB}$ difference). Overall, the gap stays within roughly $0.1$--$0.2\,\text{dB}$, which, combined with Fig.~\ref{fig:SCROOGE_EE} (up to $\sim 58\%$ elements OFF at low load and $\sim 24\%$ even at high load), demonstrates a favorable performance--energy trade-off: substantial element deactivation with essentially unchanged multiplexing behavior and preserved tier-aware resource allocation.

A closer inspection of Fig.~\ref{fig:SCROOGE_with_withoutEE} shows that, for $K=6$ and $K=8$, enabling the EE mechanism yields a slightly smaller SNR loss than leaving it inactive. This is physically plausible because elements with persistently low influence across the considered users may contribute weak, poorly aligned, or even conflicting field components under the shared multi-user configuration~\cite{6899611}, especially under coarse phase quantization. Deactivating such elements can therefore suppress marginal reflections and reduce coupling-induced perturbations, resulting in a more coherent effective aperture and, in some cases, a lower overall SNR loss~\cite{9360851}. This behavior should thus be interpreted as a combined energy-efficiency and EM regularization effect rather than as a purely power-saving operation. A deeper investigation of the precise EM conditions under which this improvement occurs is beyond the scope of the present analysis.

\subsection{Physics-aware Admission Control}
As concerns the admission control evaluation, Fig.~\ref{fig:AcceptanceRatio_AC} reports the acceptance ratio per tier and captures how SCROOGE's tier-dependent physics/QoS gating translates into admission probabilities.
The results exhibit a clear monotonic trend: acceptance increases substantially from high tiers to low tiers, consistent with the policy that higher tiers are protected by stricter phase-mismatch tolerances (smaller allowable $x\%$ of $2\pi$), whereas lower tiers are admitted more easily (larger allowable mismatch and/or relaxed stress constraints).
Numerically, the admission controller accepts $43.3\%$ of Tier~$1$ candidates ($175/404$), $49.6\%$ of Tier~$2$ ($201/405$), $61.3\%$ of Tier~$3$ ($247/403$), $83.0\%$ of Tier~$4$ ($322/388$), and $96.0\%$ of Tier~$5$ ($384/400$). 
This pattern is exactly what SCROOGE aims for in admission control: premium tiers are admitted only when the current network configuration is highly compatible with their most influential elements ensuring premium-grade performance, while lower tiers are admitted at much higher rates, leveraging the larger admissible distortion budgets.

Fig.\ref{fig:SNRLoss_AC} evaluates the admission-control mechanism by separating the predicted SNR loss into Accepted (A) and Rejected (R) outcomes for each tier, while overlaying the corresponding baseline tier loss used as the QoS reference. A clear separation is observed across all tiers: accepted candidates remain below the baseline, whereas rejected candidates lie above it, confirming that the admission logic operates as intended. The tier baselines increase progressively from Tier$1$ to Tier~$5$, namely $7.81$, $10.35$, $13.48$, $17$, and $19.92\,\mathrm{dB}$, respectively, reflecting the operator-defined QoS requirements.

More specifically, the results per tier are as follows:
\begin{itemize}
\item \textbf{Tier~$1$} ($7.81\,\mathrm{dB}$ baseline): accepted candidates achieve $\mathrm{mean}(A)=2.8\,\mathrm{dB}$ and $\max(A)=7.8\,\mathrm{dB}$, while rejected candidates yield $\mathrm{mean}(R)=14.35\,\mathrm{dB}$ and $\min(R)=7.91\,\mathrm{dB}$.
\item \textbf{Tier~$2$} ($10.35\,\mathrm{dB}$ baseline): accepted candidates achieve $\mathrm{mean}(A)=4.6\,\mathrm{dB}$ and $\max(A)=10.35\,\mathrm{dB}$, while rejected candidates yield $\mathrm{mean}(R)=15.3\,\mathrm{dB}$ and $\min(R)=10.39\,\mathrm{dB}$.
\item \textbf{Tier~$3$} ($13.48\,\mathrm{dB}$ baseline): accepted candidates achieve $\mathrm{mean}(A)=5.61\,\mathrm{dB}$ and $\max(A)=13.46\,\mathrm{dB}$, while rejected candidates yield $\mathrm{mean}(R)=17\,\mathrm{dB}$ and $\min(R)=13.49\,\mathrm{dB}$.
\item \textbf{Tier~$4$} ($17\,\mathrm{dB}$ baseline): accepted candidates achieve $\mathrm{mean}(A)=8.36\,\mathrm{dB}$ and $\max(A)=16.93\,\mathrm{dB}$, while rejected candidates yield $\mathrm{mean}(R)=20.08\,\mathrm{dB}$ and $\min(R)=17.03\,\mathrm{dB}$.
\item \textbf{Tier~$5$} ($19.92\,\mathrm{dB}$ baseline): accepted candidates achieve $\mathrm{mean}(A)=9.06\,\mathrm{dB}$ and $\max(A)=19.88\,\mathrm{dB}$, while rejected candidates yield $\mathrm{mean}(R)=22.07\,\mathrm{dB}$ and $\min(R)=20.08\,\mathrm{dB}$.
\end{itemize}

In all tiers, the maximum accepted loss remains just below the corresponding baseline, as expected from a strict QoS gate, whereas the minimum rejected loss already exceeds it. The spread of outcomes further supports the robustness of the mechanism: accepted candidates exhibit moderate variability, with $\mathrm{std}(A)=2.79$–$5.95\,\mathrm{dB}$, since some operate close to the feasibility boundary, while rejected candidates remain consistently above threshold, with $\mathrm{std}(R)=2.76$–$3.98\,\mathrm{dB}$. Overall, Fig.~\ref{fig:SNRLoss_AC} confirms that SCROOGE enforces tier-specific QoS sharply while admitting only candidates whose predicted losses remain safely below the corresponding tier baseline.

\section{Conclusion}\label{sec:conclusion}

This paper introduced SCROOGE, a unified physics-aware orchestration framework for RIS-assisted networks that jointly addresses resource allocation, energy efficiency, and admission control during the operating phase. The main premise of SCROOGE is that the products of offline physics-based codebook compilation—namely user-specific optimal entries and per-element influence information—can be directly exploited by the network orchestrator to make fast and meaningful RIS-control decisions without resorting to online optimization or iterative RIS-user control loops. In this way, SCROOGE bridges the gap between high-fidelity RIS synthesis and practical network operation.

The presented results support this design choice. In resource allocation, SCROOGE preserves the intended tier hierarchy more effectively than the non-physics-aware baseline and yields lower SNR loss across user tiers by recognizing that RIS elements do not contribute equally to user performance. In energy efficiency, SCROOGE deactivates a considerable fraction of weakly influential elements, achieving substantial potential energy savings while introducing only negligible performance loss. In admission control, SCROOGE enforces tier-dependent QoS requirements in a clear and consistent manner, separating accepted and rejected users according to the compatibility of their most critical RIS elements with the already deployed common configuration. Overall, these results show that physics-aware descriptors extracted offline can meaningfully improve operating-phase RIS orchestration under realistic hardware constraints and structured indoor propagation conditions. A natural extension of this work is to couple the proposed influence-aware orchestration mechanisms with explicit protection constraints, so that resource allocation, energy-efficient activation, and admission control are performed not only with QoS and tier awareness, but also with physical-layer security objectives in mind.

\bibliographystyle{ieeetr}
\bibliography{refs}

\end{document}